\documentclass[a4paper,11pt]{article}

\usepackage{amsfonts,amsthm,amssymb}
\usepackage{latexsym}
\usepackage{graphicx}
\usepackage{color}
\usepackage{axodraw}

\newcommand{\eps}{\varepsilon}

\newcommand{\eq}{\begin{equation}}
\newcommand{\en}{\end{equation}}
\newcommand{\eqa}{\begin{eqnarray}}
\newcommand{\ena}{\end{eqnarray}}

\newcommand{\sss}[2]{$#1$$=$$#2$}
\newcommand{\ds}[1]{$#1$}

\newcommand{\Id}{{\mathrm{Id}}}
\newcommand{\F}{\mathbb{F}}
\newcommand{\Vcal}{{\mathcal V}}
\newcommand{\Lcal}{{\mathcal L}}
\newcommand{\Scal}{{\mathbb S}}
\newcommand{\Ocal}{{\mathcal O}}
\newcommand{\Rcal}{{\mathcal R}}

\newcommand{\dx}{\!\!{\rm d}^4x\,}

\newcommand{\dk}{\!\!{\rm d}^4k\,}

\begin{document}

\thispagestyle{empty}

\vspace*{0.1cm}

\begin{center}
{\bf \LARGE  Hopf Algebraic Structures in the Cutting Rules \\[2mm] }

\vspace{1cm}

Yong Zhang${}$
\\[0.7cm]

${}$ Institute of Theoretical Physics, Chinese Academy of Sciences\\
P. O. Box 2735, Beijing 100080, P. R. China\\

\end{center}

\vspace{0.5cm}

\begin{center}
\parbox{12cm}{
\centerline{ \small  \bf Abstract }      \small \noindent

\vspace{1.0cm}

Since the Connes--Kreimer Hopf algebra was proposed, revisiting
present quantum field theory has become meaningful and important from
algebraic points. In this paper, the Hopf algebra in the cutting
rules is constructed. Its coproduct contains all necessary
ingredients for the cutting equation crucial to proving the
perturbative unitarity of the S-matrix. Its antipode is compatible
with the causality principle. It is obtained by reducing the Hopf
algebra in the largest time equation which reflects partitions of
the vertex set of a given Feynman diagram. First of all, the
Connes--Kreimer Hopf algebra in the BPHZ renormalization
instead of the dimensional regularization and the minimal
subtraction is described so that the strategy of setting up Hopf
algebraic structures of Feynman diagrams becomes clear.
 }

 \end{center}

 \vspace*{10mm}
 \begin{tabbing}

\end{tabbing}

\newpage

\section{Introduction}

 Quantum field theory combines the special relativity
 with quantum mechanics to present a theoretical description of particle
 physics. So far known experiments in high energy physics
 can be explained by the standard model which is a non-abelian
 gauge field theory with spontaneous breaking symmetries \cite{peskin}.  But
 calculating quantum corrections has to choose regularization
 and renormalization schemes to extract finite physical quantities from infinities.
 Although present quantum field theory may be considered as an effective field theory of
 more fundamental theories, we ask what is behind successful
 regularization and renormalization schemes. As Dirac \cite{dirac} argued:
``some fundamental change in our ideas'', answering
 this question is so meaningful to both fundamental physics and new mathematics.
 Recent developments in perturbative quantum field theory suggest
 that it is one possible way out to go through quantum field theory from algebraic points.

 The BPH \cite{bogo, hepp} or BPHZ \cite{zim1} renormalization proposes a general
 methodology of treating overlapping divergences of Feynman diagrams
 consistent with the locality principle. Recently,  BPH  has been found
 to be described by the Connes--Kreimer Hopf algebra \cite{kreimer1, kreimer2}.
 The axioms of the Hopf algebra are listed in the appendix. For a divergent Feynman diagram
 $\Gamma$, its coproduct generates all possible disjoint unions of divergent subdiagrams required
 by the R-operation \cite{bogo} and its twisted antipode yields the Zimmermann forest formula
 \cite{zim1} in the dimensional regularization with the minimal subtraction. More
 attractively, the Connes--Kreimer Hopf algebra is the same as the Hopf algebra of rooted
 trees \cite{kreimer3} in non-commutative geometry. It also relates the Birkhoff decomposition,
 the Riemann-Hilbert problem \cite{kreimer4} and the Baxter algebra together.  In
 this paper, we will present the Connes--Kreimer Hopf algebra in BPHZ so that the
 strategy of constructing Hopf algebras of Feynman diagrams is explicit, as guides us to obtain
 and explain the Hopf algebra in the cutting rules.

The S-Matrix in quantum field theory calculates the cross section
which is measured in scattering experiments. Its perturbative
unitarity is realized at the level of Feynman diagrams as a
diagrammatic equation which has been found to be similar to the
cutting equation from the cutting rules \cite{cutkosky}. In
\cite{veltman} the cutting equation was derived by the largest
time equation. The latter expresses one equality of Feynman
integrands by replacing Feynman propagators $\Delta_F$ with
cutting propagators $\Delta_+$ and $\Delta_{-}$ in a given Feynman
integrand. Its formalism of Feynman integrals is simplified to the
cutting equation by removing all terms violating the conservation
of energy which contain at least one vertex connecting other
vertexes only through $\Delta_+$ ($\Delta_{-}$). There is a recent
review on the largest time equation and the cutting rules 
\cite{vanniu}.

With the strategy of setting up the Connes--Kreimer Hopf algebra,
we construct the Hopf algebra in the largest time equation and
reduce it to the Hopf algebra in the cutting rules \cite{yong2}. The previous
Hopf algebra represents partitions of the vertex set of a Feynman diagram $\Gamma$.
Its antipode is the complex conjugation of the Feynman integral $\Phi(\Gamma)$.
The coproduct of the latter one leads to the cutting equation and the corresponding
antipode is an advanced function vanishing in retarded regions.

The paper is organized as follows. In section 2, the Connes--Kreimer Hopf algebra
in BPHZ instead of the dimensional regularization is presented to show a general
procedure of constructing Hopf algebras of Feynman diagrams. In section 3, the Hopf
algebra in the cutting rules is obtained by reducing the Hopf algebra in the largest
time equation. In conclusion, some remarks are made. In appendix, the axioms of the
Hopf algebra are sketched.

\section{The Connes--Kreimer Hopf algebra in BPHZ}

This section presents the Connes--Kreimer Hopf algebra in
Zimmermann's sense \cite{zim1}, in order to show a general
procedure of realizing  Hopf algebraic structures of Feynman
diagrams in perturbative quantum field theory. All necessary graph
terminologies are introduced.

A connected Feynman diagram $\Gamma$ consists of vertexes,
external lines and internal lines connecting two vertexes. In
BPHZ, it is completely fixed by $\Lcal(\Gamma)$ denoting the set
of its all lines. Let \ds{\Lcal^e(\Gamma)} denote the set of
all external lines and \ds{\Lcal^i(\Gamma)} denote the set of
all internal lines, then
$\Lcal(\Gamma)=\Lcal^e(\Gamma)\cup\Lcal^i(\Gamma)$. The symbol
$\Vcal(\Gamma)$ denotes the set of all vertexes attaching to lines
in $\Lcal(\Gamma)$. Its subdiagram \ds{\gamma} is specified by
\ds{\Lcal(\gamma)} satisfying
\ds{\Lcal(\gamma)}\ds{\subseteq}\ds{\Lcal(\Gamma)}.

Two Feynman diagrams $\gamma$ and $\gamma^{\prime}$ are disjoint
when $\Lcal^i{(\gamma)}\cap\Lcal^i{(\gamma^{\prime})}$ is the
empty set $\varnothing$.  Let $s$ represent
$(\gamma_1,\gamma_2\cdots\gamma_c)$ with every two diagrams
disjoint. It may be the empty set $\varnothing$. The set
\ds{\Scal} of all possible $s$ has the form \eq
{\mathbb S}= \{s\mid\,\,
 \Lcal^i(\gamma_j)\cap\Lcal^i(\gamma_k)=\varnothing,\,
 \Lcal(\gamma_j)\subseteq\Lcal(\Gamma),
 j\neq k;j,k=1,2\cdots c \}.
\en  The symbol $\sum_{s\in\,\Scal}^{\prime\prime}$ denotes the
summation over all disjoint connected subdiagrams except
\sss{s}{(\Gamma)} and \sss{s}{\varnothing}.

 Let $\delta$ denote the disjoint union of
connected subdiagrams $\gamma_1, \gamma_2, \cdots\gamma_c$. The
reduced subdiagram $\Gamma/\delta$ is given by contracting each
connected part of \ds{\delta} to a point so that
\ds{\Gamma/\delta} is specified by  $\Lcal(\Gamma/\delta)
=\Lcal(\Gamma)\setminus \Lcal^i(\delta)$ and $\Vcal(\Gamma/\delta)
=\Vcal(\Gamma)\setminus \Vcal(\delta)\cup \{\overline{V}_1,
\overline{V}_2\cdots\overline{V}_c\}$, where $\overline{V}_i$,
\sss{i}{1,2\cdots c,} are new vertexes from contraction. Here
\ds{A\setminus B} denotes the difference between two sets $A$ and $B$.

 In BPHZ, divergent Feynman diagrams are well treated. Let
 $I_{\Gamma}(K^{\Gamma},q^{\Gamma})$ denote the unrenormalized
 Feynman integrand of $\Gamma$, $K^{\Gamma}$ being its internal
 momenta and $q^{\Gamma}$ being its external momenta. Let $\Rcal$
 denote a renormalization map extracting a divergent part of a
 Feynman diagram $\gamma$. It has
 the form ${\cal R}^{\gamma}={\cal R}^{\gamma}(\gamma)=(-t^{\gamma})S_{\gamma}$,
 $t^{\gamma}$ being
 the Taylor subtraction operator cut off by the divergence degree
 of $\gamma$ and $S_{\gamma}$ being the substitution operator
 satisfying  \eq S_{\mu}:~~~K^{\gamma}\to K^{\gamma}(K^{\mu}),\qquad
 q^{\gamma}\to q^{\gamma}(K^{\mu},q^{\mu}),
 ~\mbox{for}~  \gamma\subset\mu.
 \en

The global divergence $\Ocal_\Gamma$ is obtained by
$\mathcal{O}_{\Gamma}=-\mathcal{R}(\overline{R}_{\Gamma})$ in
which $\overline{R}_{\Gamma}$ does not contain any subdivergences
and is calculated in a recursive way {\setlength\arraycolsep{2pt}
\eq \label{(3)}
\overline{R}_{\Gamma}=I_{\Gamma}(K^{\Gamma},q^{\Gamma})+
{\sum_{s\in\,\mathbb{S}}}^{\prime\prime}
I_{\Gamma/{\gamma_1\gamma_2\cdots\gamma_c}}(K,q)\prod_{\tau=1}^c
\mathcal{O}_{\gamma_\tau}(K^{\gamma_\tau},q^{\gamma_\tau}), \en}
Therefore the renormalized Feynman integrand $R_{\Gamma}$ has the
form \eqa
{R}_{\Gamma}&=&\overline{R}_{\Gamma}+\Ocal_{\Gamma} \nonumber\\
&=&{\sum_{s\in\,\mathbb{S}}}
I_{\Gamma/{\gamma_1\gamma_2\cdots\gamma_c}}(K,q)\prod_{\tau=1}^c
\mathcal{O}_{\gamma_\tau}(K^{\gamma_\tau},q^{\gamma_\tau}). \ena

Now we come to the Connes--Kreimer Hopf algebra
$(H,+,m,\eta,\Delta,\eps,S;\F)$ \cite{kreimer1, kreimer2}. Its all
elements are specified by \eqa \left\lbrace\begin{array}{lll}
H:&& \mbox{the set generated by 1PI Feynman diagrams};\\
+:&& \mbox{the linear combination}; \\
\F:&&\mbox{the complex number $\mathbb{C}$ with the unit 1};\\
m:&& \mbox{the disjoint union},\,\Gamma_1\Gamma_2=m(\Gamma_1\otimes\Gamma_2)=\Gamma_1\cup\Gamma_2; \\
e:&& \mbox{the empty set: $\varnothing$ and }\eta(1)=e;\\
\Delta: &&\Delta(\Gamma)=\Gamma\otimes e+e\otimes\Gamma+\nonumber\\
 &   &\sum_{s\in\,\mathbb{S}}^{\prime\prime}\gamma_1\gamma_2\cdots\gamma_c\otimes
\Gamma/{\gamma_1\gamma_2\cdots\gamma_c};\\
\eps:&& \eps(e)=e;\,\,\, \eps(\Gamma)=0, \,\, {\rm if}\,\, \Gamma\neq e;\\
S:&& S(\Gamma)=-\Gamma-\sum_{s\in\,\mathbb{S}}^{\prime\prime}S(\gamma_1\gamma_2\cdots\gamma_c)
\Gamma/{\gamma_1\gamma_2\cdots\gamma_c}.
\end{array}
\right. \ena It satisfies all the axioms of the Hopf algebra
listed in the appendix. The Feynman rules $\Phi$ can be regarded
as a map from Feynman diagrams to unrenormalized Feynman
integrands $\Phi:\,\Gamma\to \Phi(\Gamma)$ which is a character or
a nonzero homomorphism. The twisted antipode $S_R(\Gamma)$, \eq
\label{(5)} S_R(\Gamma)=-\Rcal[\Phi(\Gamma)]-
\Rcal\,\,{\sum_{{s\in\,\mathbb{S}}}}^{\prime\prime} S_R
(\gamma_1\gamma_2\cdots\gamma_c)\Phi(\Gamma/\gamma_1\gamma_2\cdots\gamma_c)
\en is also a character when the renormalization map \ds{\Rcal}
satisfies  \eq
\Rcal(\gamma_1\gamma_2)+\Rcal(\gamma_1)\Rcal(\gamma_2)
=\Rcal(\gamma_1\Rcal(\gamma_2))+\Rcal(\Rcal(\gamma_1)\gamma_2) \en
for two disjoint diagrams $\gamma_1$ and $\gamma_2$, which means
$\cal R$ is a typical Baxter operator \cite{kurusch}.

 To relate BPHZ to the Connes--Kreimer Hopf algebra, we specify $ \Ocal_{\Gamma} \Leftrightarrow
 S_R(\gamma)$, $R_\Gamma \Leftrightarrow S_R\star\Phi(\Gamma)$. The latter one is a convolution
 between two characters $\Phi$ and $S_R$. From this comparison, the coproduct includes all possible
 disjoint unions of divergent subdiagrams in $\overline{R}_{\Gamma}$  and the twisted antipode (\ref{(5)})
 is defined with the same recursive procedure as by $\overline{R}_{\Gamma}$. Hence setting up interesting
 Hopf algebraic structures is to define a coproduct or antipode satisfying the axioms of the Hopf algebra
 with physical interpretations.

\section{Hopf algebraic structures in the cutting rules}

   This section gives the Hopf algebra in the cutting rules \cite{cutkosky}. Its coproduct is directly
   related to the cutting equation \cite{cutkosky}. It is obtained by reducing the Hopf algebra  in
   the largest time equation \cite{veltman, vanniu} derived by applying the circling rules to circled diagrams.
   Relating partitions of the vertex set of a given Feynman diagram to circled diagrams, the Hopf algebra for
   the largest time equation is specified. Here only the scalar field theory $\lambda\,\varphi^4$, $\lambda$ being
   the coupling constant, is considered.

\subsection{The Hopf algebra in the largest time equation}

The largest time equation is an equality of Feynman integrands. It
reflects the decomposition of the Feynman propagator $\Delta_F$
into the sum of the positive cutting propagator $\Delta_{+}$ and
the negative cutting propagator $\Delta_{-}$: \eq
\Delta_F(x-y)=\theta(x^0-y^0)\,\Delta_{+}(x-y)+
\theta(y^0-x^0)\,\Delta_{-}(x-y), \en where the theta function is
a normal step function and $\Delta_{+}, \Delta_{-}$ have the forms \eq
\Delta_{+}(x-y)=\Delta_{-}(y-x)=\int\,{\frac {\dk} {(2\pi)^4}}\,
\theta(k_0)\, 2\pi\,\delta(k^2+m^2) \,e^{i\,k(x-y)}.\en  The
simplest largest time equation is given by \eq \label{simplest}
(\Delta_F+\Delta^\ast_F-\Delta_{+} -\Delta_{-})(x-y)=0 \en where
the conjugation propagator $\Delta^\ast_F$ is the complex
conjugation of $\Delta_F$. To obtain the largest time equation for
a general Feynman diagram, we need the circling rules for circled
diagrams instead of the Feynman rules.

A circled diagram is a Feynman diagram with some vertexes
encircled. For a given N-vertex connected Feynman diagram, $2^N$
circled diagrams are obtained by putting some circles around its
vertexes.  The circling rules are devised to map an integral to
a circled diagram. For a circled vertex $x$, include a factor $\int\,\dx\,(i\lambda)$;
for an internal line connecting vertexes $x_i$ and $x_j$, include the conjugation
propagator $\Delta_F^{\ast}(x_i-x_j)$; for an internal line
connecting a uncircled vertex $x_j$ to a circled vertex $x_i$,
include the positive cutting propagator $\Delta_{+}(x_i-x_j)$; for
an internal line connecting a circled vertex $x_j$ to a uncircled
vertex $x_i$, include a negative cutting propagator
$\Delta_{-}(x_i-x_j)$; for external lines, uncircled vertexes and
internal lines connecting two uncircled vertexes, apply the
Feynman rules. Especially, a diagram with all vertexes encircled
is called a conjugation diagram and the corresponding circling
rules are called the conjugation rules. As an example, the largest
time equation of an one-loop Feynman diagram is given by \eq
\label{exam1} (\Delta^2_F+\Delta^{\ast 2}_F-\Delta^2_{+}
-\Delta^2_{-})(x-y)=0 \en and with the circled diagrams it has a
diagrammatic representation
\begin{center}
  \begin{picture}(180,40)(0,0)
\Vertex(10,30){1.5} \Oval(20,30)(5,10)(0) \Vertex(30,30){1.5}
\Text(10,26)[tr]{$x$}  \Text(30,26)[tl]{$y$}  
\Text(40,30)[]{$+$} \CArc(50,30)(1.5,0,360)
 \Oval(60,30)(5,9)(0)\CArc(70,30)(1.5,0,360)
\Text(50,26)[tr]{$x$}  \Text(70,26)[tl]{$y$}  
\Text(80,30)[]{$+$}
 \Vertex(90,30){1.5}
 \Oval(100,30)(5,9)(0)\CArc(110,30)(1.5,0,360)
\Text(90,26)[tr]{$x$}  \Text(110,26)[tl]{$y$}
\Text(120,30)[]{$+$} \CArc(130,30)(1.5,0,360)
 \Oval(140,30)(5,9)(0)\Vertex(150,30){1.5}
 \Text(130,26)[tr]{$x$}  \Text(150,26)[tl]{$y$}
 \Text(170,30)[]{$=0$}
\end{picture}
\vspace{-.5cm}
  {\\ Figure 1. An example for the largest time equation.}
  \end{center}

With the strategy of constructing the Connes--Kreimer Hopf algebra, let us
present the Hopf algebra
in the largest time equation. The set $H$ is generated by all connected
Feynman diagrams with oriented external lines. The field $\F$ is the complex
number $\mathbb{C}$ with unit $1$. The addition $+$ is defined by the
linear combination: $a\,\Gamma_1\,+\,b\,\Gamma_2\in H $,
$a,b\in\mathbb{C},~~\Gamma_1,\Gamma_2\in H$. The multiplication $m$
of two Feynman diagrams $\Gamma_1$ and $\Gamma_2$ are their
disjoint union $ \Gamma_1\Gamma_2=: \Gamma_1\cup\Gamma_2$ which
gives the associativity axiom (1). The unit map $
{\eta}$ specifies the unit $e$ as the empty set $\varnothing$ with
\sss{\eta(1)}{e}. Hence $(H,+,m,\eta;\F)$ is an algebra.

 For a connected N-vertex Feynman diagram $\Gamma$, the set of its all
 vertexes is denoted by $\Vcal_N$. A subdiagram $\gamma$ is
 specified by a subset $\Vcal_c$ of $\Vcal_N$ and all lines connecting vertexes
 in $\Vcal_c$. A reduced subdiagram $\Gamma/\gamma$ is the remained part cutting
 $\gamma$ out of $\Gamma$. Cut internal lines in $\Gamma$ are separated
 into external lines of $\gamma$ and $\Gamma/\gamma$. $\Gamma$ and
 the empty set $\varnothing$ are called trivial subdiagrams.

 The coproduct  of $\Gamma$ is defined by
$\Delta(\Gamma)=\sum_{P}\,\gamma(\Vcal_c)\otimes\,\Gamma/\gamma$,
where $P$ denotes all  partitions of $\Vcal_N$ and the
summation is over all subdiagrams $\gamma(\Vcal_c)$. It
has a expanded form \eq \Delta(\Gamma)=\,\Gamma\otimes e+ e\otimes \Gamma\,
 \sum_{1\leq c<N}\,\gamma(\Vcal_c)\otimes\,\Gamma/\gamma. \en
 The counit $\eps$ vanishes except $\eps(e)=1$. The coassociativity
is ensured by the fact that dividing a vertex set and further
dividing the vertex set of its subdiagram is equivalent to
dividing a vertex set and further dividing the vertex set of its
reduced subdiagram. The antipode $S$ is specified by
 $S(\Gamma)=-\Gamma- \sum_{1\leq c<N}\,S(\gamma(\Vcal_c))\,\Gamma/\gamma$,
 with $S(e)=e$, which is derived from the antipode axiom $(8)$.

 The coproduct of the Feynman diagram in Fig.1 has a diagrammatic representation
\begin{center}
  \begin{picture}(230,40)(0,0)
 \Text(-5,30)[]{$\Delta ($}
 \Vertex(10,30){1.5} \Vertex(30,30){1.5}
 \Text(10,26)[rt]{$x$} \Text(30,26)[lt]{$y$}
 \Oval(20,30)(5,10)(0)  \Text(40,30)[]{$)$}  
 \Text(50,30)[]{$=$}\Vertex(60,30){1.5} \Vertex(80,30){1.5}
 \Text(60,26)[rt]{$x$} \Text(80,26)[lt]{$y$}
 \Oval(70,30)(5,10)(0)
 \Text(90,30)[]{$\otimes$} \Text(100,30)[]{$e$}    
 \Text(110,30)[]{$+$}\Text(120,30)[]{$e$}\Text(130,30)[]{$\otimes$}
 \Vertex(140,30){1.5} \Vertex(160,30){1.5}
 \Text(140,26)[rt]{$x$} \Text(160,26)[lt]{$y$}
 \Oval(150,30)(5,10)(0)      
   \Text(170,30)[]{$+$}
   \Vertex(180,30){1.5}
   \Line(180,30)(190,40)  \Line(180,30)(190,20) \Text(180,26)[tr]{$x$}
   \Text(200,30)[]{$\otimes$} \Vertex(220,30){1.5}
  \Line(220,30)(210,40) \Line(220,30)(210,20)
   \Text(220,26)[tl]{$y$}  
  \Text(230,30)[]{$+$}
 \Vertex(250,30){1.5}
  \Line(240,40)(250,30)  \Line(240,20)(250,30) \Text(250,26)[tl]{$y$}
  \Text(260,30)[]{$\otimes$} \Vertex(270,30){1.5}
  \Text(270,26)[tr]{$x$} \Line(280,20)(270,30) \Line(280,40)(270,30)
  \end{picture}
  \vspace{-.5cm}
  {\\ Figure 2. An example for the coproduct in the circling rules.}
  \end{center}
To recover the largest time equation (\ref{exam1}), we introduce a
character $\Phi_c$ denoting the conjugation rules and a new
associative multiplication $m$ representing the following
integration \eq \label{coproduct} \Delta_{+}(x-y)=\int\,{\rm
d}^3k\, {\frac {e^{-ik\,y}}{\sqrt{(2\pi)^3 2\omega_k}}}\, {\frac
{e^{ik\,x}}{\sqrt{(2\pi)^3 2\omega_k}}} \en which reflects that
the positive cutting propagator $\Delta_+$ is decomposed into
integration over the phase space of an incoming physical particle
 and an outgoing physical particle and where $\omega_k=\sqrt{k^2+m^2}$.
 Hence $\Phi\star\Phi_c(\Gamma)=\Phi_c\star\Phi(\Gamma)=0$
 gives the largest time equation. As the second example, the simplest large time equation
 (\ref{simplest}) of  the Feynman propagator $\Delta_F$ has a
 diagrammatic representation
 \begin{center}
  \begin{picture}(280,20)(0,0)
  \Text(-5,10)[]{$\Delta ($}        
  \Line(5,10)(25,10)
  \Vertex(5,10){1.5}\Vertex(25,10){1.5}
  \Text(5,8)[lt]{$x$} \Text(25,8)[rt]{$y$}
  \Text(30,10)[]{$)$}              
  \Text(40,10)[]{$=$}
  \Vertex(50,10){1.5}\Vertex(70,10){1.5}\Line(50,10)(70,10)
  \Text(50,8)[lt]{$x$} \Text(70,8)[rt]{$y$}
  \Text(80,10)[]{$\otimes$}
  \Text(90,10)[]{$e$}                  
  \Text(100,10)[]{$+$}
  \Text(110,10)[]{$e$}  \Text(120,10)[]{$\otimes$}
  \Vertex(130,10){1.5}  \Vertex(150,10){1.5}\Line(130,10)(150,10)
  \Text(130,8)[lt]{$x$} \Text(150,8)[rt]{$y$}     
  \Text(160,10)[]{$+$}
  \Vertex(170,10){1.5}\Line(170,10)(180,10)
   \Text(170,8)[lt]{$x$}
  \Text(190,10)[]{$\otimes$}
  \Vertex(210,10){1.5}\Line(200,10)(210,10)
   \Text(210,8)[rt]{$y$}                    
  \Text(220,10)[]{$+$}
  \Vertex(240,10){1.5}\Line(230,10)(240,10)
  \Text(240,8)[rt]{$y$}
  \Text(250,10)[]{$\otimes$}
  \Vertex(260,10){1.5}\Line(260,10)(270,10)
   \Text(260,8)[lt]{$x$}                      
  \end{picture}
   {\vspace{.3cm}\\ Figure 3. The coproduct of the Feynman propagator.   }
  \end{center}
 Furthermore, with the new multiplication $m$ in (\ref{coproduct}), its antipode has a
 diagrammatic form
\vspace{-.2cm}
\begin{center}
\begin{picture}(170,40)(0,0)
  \Text(0,30)[]{$S($}
  \Vertex(10,30){1.5}    \Vertex(30,30){1.5}
  \Line(10,30)(30,30)    \Text(40,30)[]{$ )$}
 \Text(10,28)[lt]{$x$} \Text(30,28)[rt]{$y$}  
  \Text(50,30)[]{$=$}    \Text(60,30)[]{$-$}
  \Line(70,30)(90,30) \Vertex(70,30){1.5}
   \Vertex(90,30){1.5}
  \Text(70,28)[lt]{$x$} \Text(90,28)[rt]{$y$}      
   \Text(100,30)[]{$+$}
    \Vertex(110,30){1.5}   \CArc(128.5,30)(1.5,0,360)
     \Line(110,30)(127,30)
    \Text(110,28)[lt]{$x$} \Text(128.5,28)[rt]{$y$}       
    \Text(140,30)[]{$+$}
    \CArc(151.5,30)(1.5,0,360)   \Vertex(170,30){1.5}
     \Line(153,30)(170,30)
 \Text(151.5,28)[lt]{$x$} \Text(170,28)[rt]{$y$}
   \end{picture}
  {\\ \vspace{-.5cm} Figure 4. The antipode of the Feynman propagator.}
  \end{center}
 Hence the antipode of $\Phi(\Gamma)$ is explained as its conjugation
 obtained  by applying the conjugation rules to the conjugation diagram
 of $\Gamma$. The convolution $\Phi\star S(\Gamma)=S\star \Phi(\Gamma)=0$ leads to the largest
 time equation.

\subsection{The Hopf algebra in the cutting rules}

The cutting propagators $\Delta_{+}$ ($\Delta_{-}$) in the
circling rules allows the positive (negative) energy to flow in lines, as
is not always consistent with the conservation of energy at each
vertex. For example, circled diagrams with one vertex relating
other vertexes only through $\Delta_{+}$ ($\Delta_{-}$) correspond
to vanishing Feynman integrals. Therefore the admissible cutting diagrams
and cutting rules are introduced to represent a simplified largest
time equation of Feynman integrals.

From a connected Feynman diagram $\Gamma$, an admissible cut diagram
is obtained by putting a cut line $l_c$ through it. The
admissible cut divides $\Gamma$ into two parts: The left part
$\gamma$ connects at least one incoming lines, while the right
part $\Gamma/\gamma$ connects at least one outgoing lines. Special
cases of only cutting incoming or outgoing external lines are also included.
The cutting rules construct integrals from
admissible cut diagrams. For the cut diagram $\gamma$, apply
Feynman rules; for $\Gamma/\gamma$, apply the conjugation rules;
for a cut internal line, include a positive cutting propagator
$\Delta_{+}(x-y)$, $y$ in $\gamma$ and $x$ in $\Gamma/\gamma$; and
for cut external lines,  no special rules. As an example, the
cutting equation of the one-loop Feynman diagram in Fig.1 has
the following diagrammatic representation
\begin{center}
  \begin{picture}(120,40)(0,0)
 \Vertex(10,30){1.5} \Oval(20,30)(5,10)(0) \Vertex(30,30){1.5}
\Text(10,26)[tr]{$x$}  \Text(30,26)[tl]{$y$}      
\DashLine(40,40)(40,20){5}
 \Text(50,30)[]{$+$}
\DashLine(60,40)(60,20){5}
 \CArc(70,30)(1.5,0,360)
 \Oval(80,30)(5,9)(0)\CArc(90,30)(1.5,0,360)
\Text(70,26)[tr]{$x$}  \Text(90,26)[tl]{$y$}      
 \Text(100,30)[]{$+$}
 \Vertex(110,30){1.5}\DashLine(120,40)(120,20){5}
 \Oval(120,30)(5,9)(0)\CArc(130,30)(1.5,0,360)
 \Text(150,30)[]{$=0$}
 \Text(110,26)[tr]{$y$}  \Text(130,26)[tl]{$x$}     
\end{picture}
\vspace{-.5cm}
  {\\ Figure 5. An example for the cutting equation at $x^0 > y^0$.}
  \end{center}

In the Hopf algebra in the cutting rules, the definitions of $H, +, m, \eta, \F$
and the notations for subdiagram, trivial subdiagram, reduced
subdiagram are the same as in the above. The coproduct
of $\Gamma$ expands as \eq \Delta(\Gamma)=\,\Gamma\,\otimes e\,+\,e\otimes \Gamma\,
 {\sum_{\mbox{\tiny all admissible cuttings}}}^{\prime\prime}\,\gamma\otimes\,
 \Gamma/\gamma,
\en where the double prime denote the summation over nontrivial
subdiagrams. As an example, the coproduct of the one-loop diagram
in Fig.1 has an diagrammatical form
\begin{center}
 \begin{picture}(170,40)(0,0)
 \Text(-5,30)[]{$\Delta ($}
 \Vertex(10,30){1.5} \Vertex(30,30){1.5}
 \Text(10,26)[rt]{$x$} \Text(30,26)[lt]{$y$}
 \Oval(20,30)(5,10)(0)  \Text(40,30)[]{$)$}  
 \Text(50,30)[]{$=$}\Vertex(60,30){1.5} \Vertex(80,30){1.5}
 \Text(60,26)[rt]{$x$} \Text(80,26)[lt]{$y$}
 \Oval(70,30)(5,10)(0)
 \Text(90,30)[]{$\otimes$} \Text(100,30)[]{$e$}    
 \Text(110,30)[]{$+$}\Text(120,30)[]{$e$}\Text(130,30)[]{$\otimes$}
 \Vertex(140,30){1.5} \Vertex(160,30){1.5}
 \Text(140,26)[rt]{$x$} \Text(160,26)[lt]{$y$}
 \Oval(150,30)(5,10)(0)      
  \Text(170,30)[]{$+$}
 \Vertex(190,30){1.5}
  \Line(180,40)(190,30)  \Line(180,20)(190,30) \Text(190,26)[tl]{$y$}
  \Text(200,30)[]{$\otimes$} \Vertex(210,30){1.5}
  \Text(210,26)[tr]{$x$} \Line(220,20)(210,30) \Line(220,40)(210,30)
  \end{picture}
  \vspace{-.5cm}
  {\\ Figure 6. An example for the coproduct in the cutting rules.}
  \end{center}

The counit $\eps$ is defined the same as before. The coassociativity axiom $(3)$ is
satisfied because all admissible cut diagrams like
\ds{\gamma_1\otimes\gamma_2\otimes\gamma_3} are obtained by cutting $\Gamma$ twice,
and the order of two cuttings is irrelevant. The antipode $S$ is specified by $S(\Gamma)=-\Gamma-
{\sum_{\mbox{\tiny all admissible cuttings}}}^{\prime\prime}
\,S(\gamma)\,\Gamma/\gamma$ with $S(e)=e$. The cutting equation is denoted by
$m(\Phi\otimes\Phi_c)\Delta(\Gamma)=0$, where the multiplication $m$ defined by (\ref{coproduct})
has a diagrammatic form
\begin{center}
  \begin{picture}(100,30)(0,0)
  \Text(0,20)[]{$m($}
  \Vertex(20,20){1.5} \Line(10,20)(20,20)
  \Text(30,20)[]{$\otimes$}
 \Vertex(40,20){1.5} \Line(40,20)(50,20)
   \Text(60,20)[]{$ )$}
 \Text(40,18)[lt]{$x$} \Text(20,18)[rt]{$y$}
  \Text(70,20)[]{$=$}  
  \Line(83,20)(100,20)
   \Vertex(100,20){1.5} \CArc(81.5,20)(1.5,0,360)
  \Text(81.5,18)[lt]{$x$} \Text(100,18)[rt]{$y$}        
   \end{picture}
  \vspace{-.3cm}
  {\\ Figure 7. The multiplication $m$ representing (\ref{coproduct}) at $x^0 > y^0$.}
  \end{center}

 The antipode of the Feynman propagator $\Delta_F(x-y)$ has a diagrammatic representation
\begin{center}
  \begin{picture}(170,40)(0,0)
  \Text(0,30)[]{$S($}
  \Vertex(10,30){1.5}    \Vertex(30,30){1.5}
  \Line(10,30)(30,30)    \Text(40,30)[]{$ )$}
 \Text(10,28)[lt]{$x$} \Text(30,28)[rt]{$y$}         
  \Text(50,30)[]{$=$}    \Text(60,30)[]{$-$}
  \Line(70,30)(90,30) \Vertex(70,30){1.5}
   \Vertex(90,30){1.5}
  \Text(70,28)[lt]{$x$} \Text(90,28)[rt]{$y$}        
    \Text(100,30)[]{$+$}
    \CArc(111.5,30)(1.5,0,360)   \Vertex(130,30){1.5}
    \Line(113,30)(130,30)
 \Text(111.5,28)[lt]{$x$} \Text(130,28)[rt]{$y$}
   \end{picture}
   \vspace{-.5cm}
  {\\ Figure 8. The antipode of the Feynman propagator at $x^0 > y^0$.}
  \end{center}
Its form of Feynman integrands is given by \eq
S(\Delta_F(x))=-\Delta_F(x)+\Delta_+(x)=\theta(-x^0)(\Delta_+(x)-\Delta_-(x))
\en
which is the advanced propagator at $y=0$. As a
generalization, the antipode of one Feynman integral in the
cutting rules is an advanced function. This reflects that the
cutting rules are compatible with the causality principle.

\section{Concluding remarks}

To conclude, we would like to emphasize again that it is so
heuristic and worthwhile to revisit present quantum field theories
from algebraic points. Besides the Connes--Kreimer Hopf algebra,
the product with Laplace parings has been found to generate the
Wick normal-ordering so that the Wick theorem has a Hopf algebraic
origin \cite{fauser1, fauser2}. In this paper, the Connes--Kreimer
Hopf algebra in BPHZ is described with well-defined graph
terminologies in order to make a rough guide of how to construct
Hopf algebraic structures of Feynman diagrams. With that, the Hopf
algebra hiding the largest time equation is obtained and
it is a realization of the complete partition of
set. Its antipode is explained as the conjugation of a given
Feynman integrand. With the conservation of energy at each vertex,
it has a simplified formalism of Feynman integrals which gives the
Hopf algebra in the cutting rules. The corresponding coproduct
carries all information for the cutting equation. More
interesting, its antipode is compatible with the causality
principle and vanishes in retarded regions.

The Hopf algebra in the cutting rules is not sensible to the
Connes--Kreimer Hopf algebra \cite{yong2}. The admissible cut in the cutting
rules is required by the conservation of energy, while the
admissible cut in the Connes--Kreimer is to disentangle
overlapping divergences. Hence the coproduct defined here are
compatible with renormalization procedures at least the
dimensional regularization and the minimal subtraction scheme.
Furthermore, reading the Hopf algebra in the cutting rules is
meaningful from the point proving perturbative unitarity of the
S-matrix: The coproduct or tensor structures hidden in the cutting
rules are crucial \cite{yong2}. Moreover, in gauge field theories
\cite{vanniu}, the cutting equation contains non-physical degrees
of freedom.  They have to be removed by applying the Ward identities or the
Slavnov--Taylor identities since the perturbative unitarity of the S-matrix
only involves physical states. The coproduct has to be defined for the
set of relevant Feynman diagrams.

\vspace{1cm}

\begin{center}
\section*{Acknowledgments}
\end{center}

 We thank P. van Nieuwenhuizen for his helpful lectures on the
 cutting rules in the ``Mitteldeutsche Physik-Combo'' and thank
 D. Kreimer, X.Y. Li and K. Sibold for helpful comments and J.P Ma
 for helpful discussions.

\appendix

\vspace{1cm}

\section{The axioms of the Hopf algebra}

Let $(H,+,m,\eta,\Delta,\eps,S;\F)$ be a Hopf algebra over $\F$.
We denote the set by $H$, the field by $\F$, the addition by $+$,
the product by $m$, the unit map by $\eta$, the coproduct by
$\Delta$, the counit by $\eps$, the antipode by $S$, the identity
map by $\Id$, and the tensor product by $\otimes$. Then the vector
space is given in terms of the triple $(H,+;\F)$ where the field
$\F$ with unit $1$. The Hopf algebra has to satisfy the following
seven axioms:

\eq \left\lbrace\begin{array}{lll}
(1)& & m(m\otimes{\Id})=m({\Id}\otimes m),\\
(2)& & m({\Id}\otimes\eta)={\Id}=m(\eta\otimes {\Id}),\\
(3)& &(\Delta\otimes{\Id})\Delta=({\Id}\otimes\Delta)\Delta,\\
(4)& & ({\Id}\otimes \epsilon)\Delta={\Id}=(\epsilon\otimes{\Id})\Delta,\\
(5)& &\Delta(ab)=\Delta(a)\Delta(b),\nonumber\\
& &\Delta(e)=e\otimes e,~~
a, b\in H,\\
(6)& & \eps(ab)=\eps(a)\eps(b),
~~\eps(e)=1,~~a,b\in H, \\
(7)& & m(S\otimes{\Id})\Delta=\eta\circ\epsilon=m({\Id}\otimes
S)\Delta.
\end{array}
\right. \en

 In (1) and (2), $m: H\otimes H\to H$, $m(a\otimes b)=ab,\, a,\, b\in H$,
 $\eta: {\F}\to H$, which denote the associative product $m$ and the linear unit map
$\eta$ in the algebra $(H,+,m,\eta;\F)$ over $\F$ respectively. In
(3) and (4), $ \Delta: H \to H\otimes H$, $\eps: H\to {\F}$, which
denote the coassociative coproduct $\Delta$ and the linear counit
map $\eps$ in the coalgebra $(H,+,\Delta,\eps;\F)$ over $\F$ and
where we used \sss{\F\otimes H}{H} or \sss{H\otimes \F}{H}. (5)
and (6) are the compatibility conditions between the algebra and
the coalgebra in the bialgebra $(H,+,m,\eta,\Delta,\eps;\F)$ over
$\F$, claiming that the coproduct $\Delta$ and the counit
$\epsilon$ are homomorphisms of the algebra $(H,+,m,\eta;\F)$ over
$\F$ with the unit $e$. (7) is the antipode axiom that can be used
to define the antipode.

 The character $\Phi$ is a nonzero linear
functional over the algebra and is a homomorphism satisfying
\sss{\Phi(ab)}{\Phi(a)\Phi(b)}. The convolution between two
characters $\Phi$ and $\Phi_c$ is defined by $\Phi\star \Phi_c:=
m(\Phi\otimes \Phi_c)\Delta$.

\end{document}